\documentclass{aa}
\usepackage{times}
\usepackage{natbib}
\usepackage[dvips]{graphicx}
\usepackage{psfig,longtable,lscape}
\usepackage{psfig}
\usepackage{rotating}
\bibpunct{(}{)}{;}{a}{}{,}

\def\RX{RX J0146.9+6121}
\def\XMM{{\em XMM--Newton}}
\def\EPIC{{\em EPIC}}
\def\MOS{{\em MOS}}
\def\Mone{{\em MOS1}}
\def\Mtwo{{\em MOS2}}
\def\pn{{\em pn}}
\def\SAX{{\em BeppoSAX}}
\def\XTE{{\em RossiXTE}}

\def\EXOSAT{{\em EXOSAT}}
\newcommand{\rxj}{RX J0059.2--7138}
\newcommand{\xtej}{XTE J0111.2--7317}
\newcommand{\fouru}{4U 1626--67}
\newcommand{\aofive}{A 0538--66}
\newcommand{\exo}{EXO 053109--6609.2}
\newcommand{\nodata}{...}

\begin{document}

\title{\XMM~observation of the Be/neutron star system \RX: a soft X--ray excess
in a low luminosity accreting pulsar\thanks{This work is based on
observations obtained with {\em XMM--Newton}, an ESA science
mission with instruments and contributions directly funded by ESA
Member States and NASA.}}

\author{N. La Palombara \& S. Mereghetti}

\institute{{INAF - Istituto di Astrofisica Spaziale e Fisica Cosmica Milano,
via Bassini 15, I-20133 Milano, Italy}}

\offprints{N. La Palombara, nicola@iasf-milano.inaf.it}

\authorrunning{N. La Palombara \& S. Mereghetti}

\titlerunning{\XMM~observation of \RX}

\abstract{We report on the \XMM~observation of the
Be/neutron star X--ray binary system \RX, a long period ($\sim$ 23 m)
pulsar in the NGC 663 open cluster. The X--ray luminosity decreased by a factor two
compared to the last observation carried out in 1998, reaching
a level of $\sim1\times10^{34}$ erg s$^{-1}$, the lowest ever
observed in this source.  The spectral analysis reveals the
presence of a significant excess at low energies over the main
power--law spectral component. The soft excess can be described by a
black--body spectrum with a temperature of about 1 keV and an
emitting region with a radius of $\sim$ 140 m. Although the current data do
not permit to ascertain whether the soft excess is pulsed or not,
its properties are consistent with emission from the
neutron star polar cap.
%the and is present at all pulse phases.
% This type of excess is rather
%common in most luminous binary pulsars but has never been detected
%before at such a low luminosity level. Therefore
% Soft spectral components have been observed in several accreting
% pulsars with luminosity above $10^{35}$ erg s$^{-1}$.
This is the third detection of a soft excess in a low
luminosity ($\sim1\times10^{34}$ erg s$^{-1}$) pulsar, the other
being X Per and 3A 0535+262, suggesting that such spectral component, observed up
to date in higher luminosity systems, is a rather common feature
of accreting X--ray pulsars. The results on these three sources indicate that, in low luminosity systems, the soft excess tends to have a higher temperature and a smaller
surface area than in the high luminosity ones.
}

\maketitle

\section{Introduction}

High Mass X--Ray Binaries (HMXRBs) are binary systems consisting of
a neutron star (NS) or, less frequently, a black hole, accreting
matter from a high mass early type star. Based on the nature of
the mass donor star, either a supergiant of O/B spectral type or a
Be star, HMXRBs can be divided in two subgroups showing different
variability properties. HMXRBs with supergiant companions tend to be persistent sources, although some of them display high and low states with X--ray flux
varying by large factor ($\sim$ 100 or more). On the other hand, neutron star Be X-ray binaries (BeXRBs) are generally transient sources, owing to the long term variability of
the equatorial discs surrounding Be stars and/or the orbital
eccentricity. For both subgroups, when the compact object is a NS,
the X--ray spectra are well described by a rather flat power--law
between 0.1 and 10 keV (photon index $\sim$1) followed by a
high--energy cutoff.

With the advent of imaging X--ray satellites a large number of
BeXRB systems has been discovered in the Small Magellanic Cloud
\citep{Haberl&Sasaki00,Israel+00,Yokogawa+03,Macomb+03,Sasaki+03,Haberl&Pietsch04}.
The observation of these sources, unaffected by the high
interstellar absorption present in the Galactic plane, makes it
possible to study in detail their X--ray spectra extending down to
energies of a few hundred eV. This has allowed to discover that most
of them have a marked soft excess above the power--law model
\citep{Nagase02,Haberl&Pietsch05}.

%Recently, \citep{Hickox+04} have proved that this is a very common
%if not ubiquitous feature intrinsic to X--ray pulsars, whose
%origin is strongly correlated with the total intrinsic source
%luminosity.

\RX\ is a BeXRB hosting a neutron star characterized by a
rotational period of about 23 minutes, among the longest observed
in X-ray pulsars. The pulsations were discovered with non-imaging
instruments on board EXOSAT and initially attributed to a
different nearby source \citep{White+87}. Subsequent observations
clarified the picture \citep{Mereghetti+93, Hellier94}
%, \ASCA~\citep{Haberl+98a},
%\XTE~\citep{Haberl+98b} and \SAX~\citep{Mereghetti+00}.
and led to the optical identification of \RX\ with the B0 IIIe
star LS I +61$^{\circ}$ 235 \citep{Coe+93}. This star is a member
of the open cluster NGC 663 \citep{Tapia+91,Fabregat+96,Pigulski+01} for which a distance
in the range between 2 and 2.5 kpc has been derived \citep{Kharchenko+05,Pandey+05}. In the
following we adopt d = 2.5 kpc.
%It is characterized by very long spin
%period of $\sim$ 25 m, i.e. the longest known period of a Be/NS
%X--ray pulsar, and in the past years was observed by \EXOSAT~
%\citep{White+87,Mereghetti+93}, \ROSAT~\citep{Hellier94},
%\ASCA~\citep{Haberl+98a}, \XTE~\citep{Haberl+98b} and
%\SAX~\citep{Mereghetti+00}.
Thanks to its relatively small distance, \RX\ is not much absorbed
and is therefore a good target to investigate the properties of
the soft X-ray emission in a Galactic source. Here we present the
results of a recent observation obtained with the \XMM\ satellite,
providing the most sensitive observation of this source ever
obtained below 2 keV.

\section{Observations and data reduction}\label{sec:2}

\RX~was observed by \XMM~between 22:40 UT of 2004 January 14 and
10:20 UT of 2004 January 15. Since the main target was the NGC 663
open cluster, \RX\  was detected at an off-axis angle of 9.3$'$.
% and the observation lasted about 42 ks.
All the three \EPIC~instruments, i.e. the \pn~camera
\citep{Struder2001} and the \Mone~and \Mtwo~cameras
\citep{Turner2001}, were active and operated in {\em Full Frame}
mode. For all of them, the Medium thickness filter was used.

We used  version 6.1 of the \XMM~{\em Science Analysis System}
({\em SAS}) to process the event files.  After the standard
pipeline processing of the data, we looked for possible periods of
high instrumental background, due to flares of soft protons with
energies less than a few hundred keV. We found that the first
$\sim$ 15 ks of the observation were affected by a soft--proton
contamination. However, since \RX~was detected with a count--rate
($\sim$ 1 cts s$^{-1}$) much higher than that of the background, the
soft--protons during the bad time intervals have a negligible
effect on the source spectral and timing analysis (their count rate in the source extraction area is less than 0.01 cts s$^{-1}$). Hence we used
for our analysis the data of the whole observation, corresponding
to exposure times of 35.4 and 41.2 ks in the \pn\ and \MOS\ cameras, respectively.

\section{Timing analysis}\label{timing}

% In  order to study the temporal behavior of \RX, for each
% instrument we extracted from the source and the background
% circular regions two different event lists. Then the times of the
% selected events were converted to the Solar System barycenter with
% the \textit{Reconstructed Orbit Files} provided by the
% \XMM~\textit{Survey Science Center}. These events were still
% tagged with discrete arrival times, corresponding to the readout
% times of the individuall CCD frames. Therefore we "randomized"
% these times by subtracting from their values a random value
% between 0 and the relevant frame time, i.e. 2.6 s and 73 ms for
% the two \MOS~and the \pn~camera respectively.

\begin{figure}[h]
\centering
\resizebox{\hsize}{!}{\includegraphics[angle=-90,clip=true]{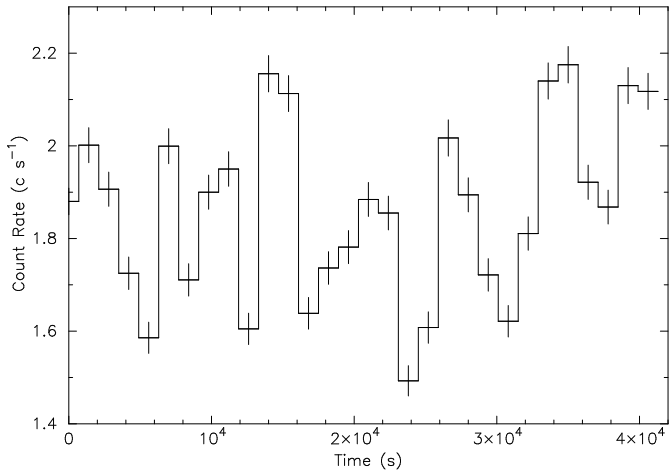}}
\caption{Background subtracted light curve  of \RX~in the energy
range 0.15--10 keV, with a time bin of 1.4 ks (i.e. 1 pulse period).}
\label{lc_1p}
\end{figure}

During our observation some flux changes on $\sim$ hour timescale were present, as shown by the background subtracted light curve plotted in Fig.~\ref{lc_1p}.
%shows that during our observation \RX varied
%in the energy range 0.15--10 keV with a bin
%time of 4.2 ks, merging the data from the three EPIC cameras.
% (the background light curves
% were renormalized to take into account the different extraction
% areas). We considered a bin time of 4.2 ks, corresponding to about
% three periods. Finally, we summed the light curves of the three
% instruments and obtained the total light curve (Fig.~\ref{lc_3p}).
%It shows that the source is not constant, since it varies up to
This is based on  the data in the 0.15--10 keV range obtained from
the three EPIC cameras. A bin size of 1.4 ks, corresponding to
one spin period, has been chosen to avoid the effects
due to the periodic pulsations. Variations up to $\sim$ 20 \%
around the average level of $\sim$ 1.9 cts s$^{-1}$ are evident.

\begin{figure}[h]
\centering
\resizebox{\hsize}{!}{\includegraphics[angle=-90,clip=true]{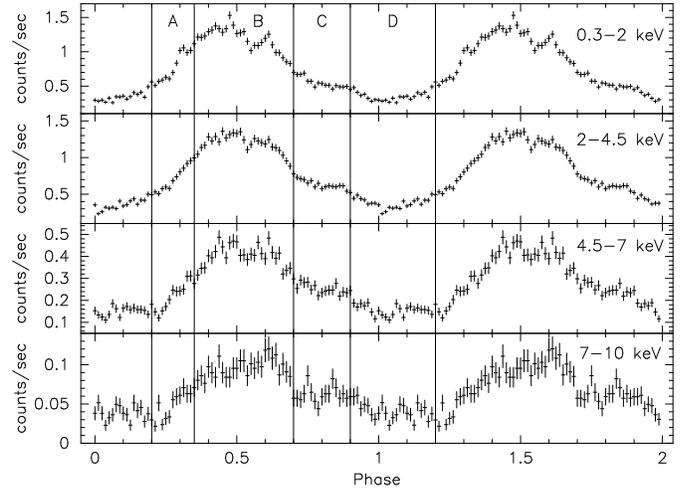}}
\caption{Background subtracted light curves of \RX~in the energy
ranges 0.3--2, 2--4.5, 4.5--7 and 7--10 keV, folded at the
best--fit period of 1396.14 s. The vertical lines indicate the phase intervals used for the spectral analysis.}\label{4LC}
\end{figure}

To obtain a  measure of the pulse period, we converted the times
of arrival to the solar system barycenter and performed a folding
analysis using the source events of three  cameras. By fitting the
$\chi^{2}$ versus trial period curve with the appropriate sinc
function as described in  \citet{Leahy1987}, we derived a period
of 1396.14 $\pm$ 0.25 s. In Fig.~\ref{4LC} we show the folded light
curves in four energy intervals (0.3--2, 2--4.5, 4.5--7 and 7--10
keV) chosen to allow a direct comparison with the light curves
previously derived  with \XTE~data \citep{Mereghetti+00}.
% using the
%\XTE~data; the last one covers only partly the hardest range
%considered by \XTE, since over 10 keV the \XMM~effective area
%becomes very low; finally, the first one samples the soft part of
%the source spectrum and was not detected by \XTE. In
%Fig.~\ref{4LC} we report the light--curves of \RX~in these energy
%ranges, folded at the period P = 1395.4 s.
The pulse profile, characterized by a broad peak, is clearly
energy-dependent, with the maximum shifting from phase
$\sim$ 0.4--0.5 at E $<$ 2 keV to phase $\sim$ 0.6 at E $>$ 7 keV.
%light curve, characterized by two broad peaks, is clearly
% energy-dependent.
% The first peak at phase $\sim$ 0.5 and the
%second one at phase $\sim$ 0.6. However, t
%The relative height of the two peaks varies as a function of
%energies: the main peak is the first one below 2 keV and the
%second one above 7 keV.

\begin{figure}[h]
\centering
\resizebox{\hsize}{!}{\includegraphics[angle=-90,clip=true]{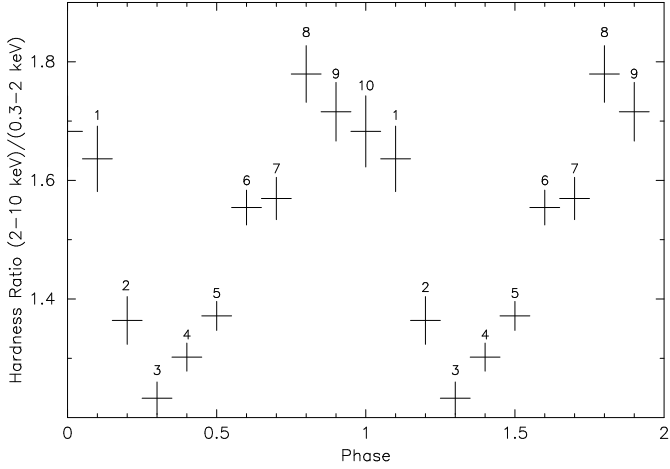}}
\caption{Hardness ratio of the two light curves of \RX~in the energy ranges 0.3-2 and 2-10 keV, as a function of the pulse phase. The HR values are obtained by folding the light curves at the best--fit period of 1396.14 s. The pulse period is divided in 10 phase bins, which are identified by progressive numbers.}\label{HR}
\end{figure}

The hardness ratio (HR) between the light curves above and below 2
keV, reported in Fig.~\ref{HR}, is characterized by a
slow and regular increase from its minimum to its maximum value,
followed by a rather irregular decrease. This plot, where we have
divided the pulse period in ten phase intervals, shows that
there is not a simple correlation between the hardness and the
total count rate:
% both the
%minimum and maximum HR values correspond to an intermediate count
%rate level; the HR continues to increase at the count rate
%turn--over around its maximum value; on the other hand, it rapidly
%decreases after the CR turn--over around its minimum value.
we observe the same HR value at completely different count
rate levels, but also very different HR values for the same count
rate. This indicates that the spectral hardness of \RX~does not
depend in a simple way on its flux level.

\section{Spectral analysis}\label{sec:3}

For the source spectra we used an extraction radius of 30$''$
around the source position in the case of the \Mtwo~and
\pn~cameras; for the \Mone~camera the extraction radius was
reduced to 15$''$, since the source was imaged close to a CCD gap.
We checked with the \textit{SAS} task \textit{epatplot} that no
event pile--up affected our data, then we accumulated all the
events with pattern range 0--4 (i.e. mono-- and bi--pixel events)
and 0--12 (i.e. from 1 to 4 pixel events) for the \pn~and the two
\MOS~cameras, respectively.  The background spectra were
accumulated on large circular areas with no sources and radius of
210$''$ and 120$''$ for the \Mone~and the \Mtwo~camera,
respectively. For the \pn~ camera the background spectrum was
extracted from a circular region of the same radius and at the
same CCD rows of the source region. We generated {\em ad hoc}
response matrices and ancillary files using the {\em SAS} tasks
{\em rmfgen} and {\em arfgen}. All spectra were rebinned with a
minimum of 30 counts per bin and fitted in the energy range
0.3--10 keV using {\em XSPEC} 11.3.2.

\begin{figure}[h]
\centering
\resizebox{\hsize}{!}{\includegraphics[angle=-90,clip=true]{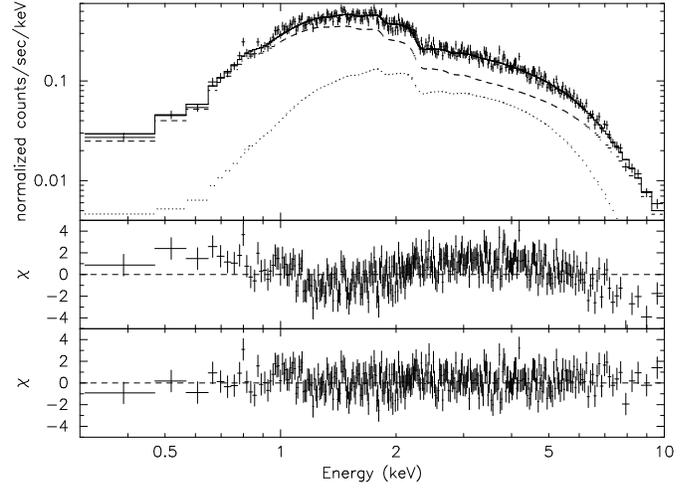}}
\caption{\textit{Top panel}: \pn\ spectrum of \RX~with the best--fit power--law (\textit{dashed line}) plus black--body (\textit{dotted line}) model. \textit{Middle
panel}: residuals (in units of $\sigma$) between data and model in the case of the single power--law. \textit{Bottom panel}: residuals in the case of the power--law plus black--body.}\label{all_spetrum_powbb}
\end{figure}

After checking that separate fits of the three spectra gave
consistent results, we fitted the spectra from the three cameras
simultaneously in order to increase the count statistics and to
reduce the uncertainties.
%We forced common  parameters for the
%three spectra and allowed only for a cross--normalization factor,
%to account for the different instrument efficiency *** ??? ***.
The fit with an absorbed power--law yielded N$_{\rm H} =
(6.24\pm0.14)\times10^{21}$ cm$^{-2}$ and photon index
$\Gamma=1.36\pm0.02$, but with large residuals and
$\chi^{2}_{\nu}$/d.o.f. = 1.283/1480\footnote{Errors are at 90 \%
confidence level for a single interesting parameter}. The addition
of a blackbody component  improved  the fit quality significantly
(Fig.~\ref{all_spetrum_powbb}): we obtained N$_{\rm H} =
(5.09^{+0.24}_{-0.23})\times10^{21}$ cm$^{-2}$,
$\Gamma=1.34^{+0.05}_{-0.06}$ and kT$_{\rm BB} =
1.11^{+0.07}_{-0.06}$ keV, with $\chi^{2}_{\nu}$/d.o.f. =
1.036/1478. The emission surface of the
thermal component has a radius R$_{\rm BB} = 140^{+17}_{-14}$ m
(for d = 2.5 kpc). The unabsorbed flux in the energy range 0.3--10
keV is $f_{\rm X}\sim2\times10^{-11}$ erg cm$^{-2}$ s$^{-1}$,
about 24 \% of which is due to the blackbody component.

We also attempted to fit the soft part of the spectra with other
emission models, such as \textit{mekal}, thermal bremsstrahlung
and broken power--law. In all these cases the results were worse
than those obtained with the blackbody model, since the
reduced chi--squared was higher, the residuals were
larger and/or the best--fit parameter values were unrealistic.

We did not find significant evidence for emission lines. By adding
Gaussian components at various energies to the above model, we
estimated upper limits of the order of 150 eV for the equivalent
width of lines in the 6--7 keV energy range.

% With the aim to assess
%also the possible presence of emission lines, we added a few
%gaussian components to the previous model at various energies. For
%the Iron line, we detected a feature at E = $7.05^{+0.11}_{-0.15}$
%keV, which is significant at a 90 \% confidence level: but it has
%$\sigma=0.06^{+0.18}_{-0.06}$ keV and an Equivalent Width of only
%38 eV. As a further test, we also considered four different fixed
%energies between 6 and 7 keV, with $\sigma$ between 0 and 0.5 keV.
%None of them was significant at 99 \% c.l. and the upper limit on
%the line Equivalent Width was always below $\sim$ 0.15 keV. We
%also found other four emission features at 0.8, 1, 2.3 and 4.2
%keV, respectively, which are significant at 90 \% c.l. at least.
%But their EQWs are lower than 25 eV, therefore we will ignore them
%in the next analysis.

\section{Phase--resolved spectroscopy}\label{spectroscopy}

Prompted by the results described above we performed a
phase--resolved spectroscopy in order to study in more detail the
source behavior. We defined phase 0 by matching the folded light
curve to that observed in 1996 with \XTE\, and extracted the
background subtracted spectra for the same four phase intervals
used in \citet{Mereghetti+00} and indicated in Fig.\ref{4LC}.
%The light curves reported in Fig.~\ref{4LC} show that the source
%variability pattern is the same one observed by \XTE, therefore we
%decided to divide the source period in the same 4 phase intervals:
%To this aim, for
%each of the three instruments first we calculated the right phase
%of all the events and, then, we extracted the
%background--subtracted spectra corresponding to each phase
%interval.
The first step was to fit all of them with the best--fit power
law plus blackbody  model of the phase averaged spectrum, leaving
only the relative normalization factors free to vary. The ratios
of the four spectra of each instrument to these renormalized
average models show significant residuals and clearly
demonstrate the spectral variability as a function of the pulse
phase.
%The comparison of these data with those reported in Fig.4 of
%\citet{Mereghetti+00} suggests a few comments, also thanks to the
%extension of the sampled energy range below 2 keV:
% This figure clearly shows the spectral variations, and in
%particular that:

%\begin{itemize}

%\item  there is a significant softening during phase interval A,
%as already observed by \XTE

%\item contrary to the previous findings, it seems that during
%phase interval C there is a spectral hardening;

%\item  during phase interval D the spectrum is harder than the
%average one above 2 keV, but below 1 keV there is also a
%significant excess that was not visible with \XTE

%\end{itemize}

We then fitted the four spectra independently. In all cases, the
absorbed power--law model was not satisfactory, while the addition
of a black--body component significantly improved the fit.
Therefore we used this model for all the spectra, leaving all the
parameters free to vary: the results are reported in
Tab.~\ref{4spectra_fit}.

% They show that, in all the four phase
% intervals, both N$_{\rm H}$ and $\Gamma_{\rm PL}$ have smaller
% values than in the corresponding \XTE~measurements. However, it is
% interesting to note that, for both of them, the variation pattern
% between the various phases is the same observed by \XTE: the
% photon index $\Gamma_{\rm PL}$ continuously decreases from phase
% intervals A to D; the absorption column N$_{\rm H}$ first
% increases from A to B and then decreases from B to D. The
% differences in the blackbody temperatures between the phase
% intervals A--B and C--D are always within the estimated errors. In
% all the phase intervals the flux fraction due to the thermal
% component is $\sim$ 25\%, i.e. comparable to the percentage
% measured in the total spectrum; only in the case of phase interval
% C it increases up to $\sim$ 40\%.

\begin{table}[htbp]
\caption{Best--fit spectral parameters for the phase--resolved spectroscopy of \RX, in the case of the independent fit of the four spectra. Errors are at 90 \% confidence level for a single interesting parameter. N$_{\rm H}$ and kT$_{\rm BB}$ are measured in units of $10^{21}$ cm$^{-2}$ and keV, respectively.}\label{4spectra_fit}
\begin{tabular}{c|cccc} \hline
Spectral &   \multicolumn{4}{c}{Phase Interval}                                          \\
Parameter &   A           &   B           &   C           &   D           \\ \hline
N$_{\rm H}$&   $4.8^{+0.4}_{-0.7}$ &   $5.4^{+0.4}_{-0.3}$ &   $5.0^{+0.8}_{-0.7}$ &   $3.6^{+0.6}_{-0.5}$ \\
$\Gamma$           &   $1.47^{+0.19}_{-0.20}$  &   $1.39^{+0.09}_{-0.08}$  &   $1.31^{+0.35}_{-0.23}$  &   $1.06^{+0.11}_{-0.21}$  \\
kT$_{\rm BB}$         &   $0.98^{+0.08}_{-0.12}$  &   $1.05^{+0.10}_{-0.08}$  &   $1.34^{+0.17}_{-0.18}$  &   $1.21^{+0.21}_{-0.16}$  \\ \hline
$f_{\rm TOT}^{a}$       &   1.81        &   3.22        &   1.74        &   1.00        \\
$f_{\rm PL}^{a}$        &   1.34    &   2.52    &   1.10    &   0.70     \\
 & (74 \%) & (78 \%) & (63 \%) & (70 \%) \\
$f_{\rm BB}^{a}$        &   0.47    &   0.70    &   0.64    &   0.30     \\
 & (26 \%) & (22 \%) & (37 \%) & (30 \%) \\ \hline
$\chi^{2}_{\nu}$/d.o.f.     &   0.972/308       &   1.021/992       &   1.085/352       &   1.131/328       \\ \hline
\end{tabular}
\begin{small}
\\
$^{a}$ Unabsorbed flux in the energy range 0.3--10 keV, in units of $10^{-11}$ erg cm$^{-2}$ s$^{-1}$
\end{small}
\end{table}

In order to investigate the relative variations  of the two
components with the period phase, we also fitted simultaneously
the four spectra forcing common values for N$_{\rm H}$, $\Gamma$ and kT$_{\rm BB}$.
In this case we obtained N$_{\rm H} = (5.8\pm0.1)\times10^{21}$ cm$^{-2}$,
$\Gamma_{\rm PL}=1.60\pm0.02$ and kT$_{\rm BB} = 1.36^{+0.04}_{-0.03}$
keV, with $\chi^{2}_{\nu}$/d.o.f. = 1.117/1989; the corresponding
normalization values are reported in Tab.~\ref{4spectra_common}.
In this interpretation the spectral changes as a function of the
phase are reproduced by the variations in the relative
contribution of the two components.

\begin{table}[htbp]
\caption{Best--fit values for the black--body and power--law normalizations, when the four spectra are fitted simultaneously with common values of N$_{\rm H}$ ($(5.8\pm0.1)\times10^{21}$ cm$^{-2}$), $\Gamma$ ($1.60\pm0.02$) and kT$_{\rm BB}$ ($1.36^{+0.04}_{-0.03}$ keV). Errors are at 90 \% confidence level for a single interesting parameter}\label{4spectra_common}
\begin{tabular}{c|cccc} \hline
Spectral            &   \multicolumn{4}{c}{Phase Interval}                                          \\
Parameter           &   A                   &   B                   &   C                   &   D           \\ \hline
R$_{\rm BB}^{a}$    &   $76\pm3$        &   $130\pm2$           &   $121\pm2$           &   $87\pm2$  \\
I$_{\rm PL}^{b}$    &   $2.24^{+0.07}_{-0.03}$  &   $3.25^{+0.05}_{-0.03}$       &   $1.32^{+0.03}_{-0.06}$      &   $0.89^{+0.01}_{-0.03}$  \\
$f_{\rm TOT}^{c}$   &   2.04                &   3.42                &   1.77                &   1.06        \\
$f_{\rm PL}^{c}$    &   1.69        &   2.46            &   1.00            &   0.67     \\
 & (83 \%) & (72 \%) & (56 \%) & (63 \%) \\
$f_{\rm BB}^{c}$    &   0.35        &   0.96            &   0.77            &   0.39     \\
 & (17 \%) & (28 \%) & (44 \%) &(37 \%) \\ \hline
\end{tabular}
\begin{small}
\\
$^{a}$ Radius of the blackbody component (in metres) for a source distance of 2.5 kpc.

$^{b}$ Intensity of the power--law component in units of $10^{-3}$ ph cm$^{-2}$ s$^{-1}$ keV$^{-1}$ at 1 keV

$^{c}$ Unabsorbed flux in the energy range 0.3--10 keV, in units of $10^{-11}$ erg cm$^{-2}$ s$^{-1}$
\end{small}
\end{table}
%However, based  on the above results we can not infer that the
%source thermal component varies as a function of the rotational
%phase. To this aim, we should prove that a constant black--body
%component is rejected by the data.
%, since in this case it would
%not be possible to obtain an acceptable fit.
%Therefore, we modified the test model by imposing a common value
%of both kT$_{\rm BB}$ and R$_{\rm BB}$ for the four spectra, while
%both $\Gamma$ and I$_{\rm PL}$ could vary within them.
%The resulting best--fit has N$_{\rm H}=
%(6.1^{+0.3}_{-0.1})\times10^{21}$ cm$^{-2}$, kT$_{\rm BB} =
%1.40^{+0.02}_{-0.04}$ keV and R$_{\rm BB}$ =
%$111^{+1}_{-4}$ m, while the power--law parameters are shown
%in Tab.~\ref{4spectra_BBcommon}. The fit quality is very good
%($\chi^{2}_{\nu}$/d.o.f. = 1.097/1989). Moreover, in this case we
%find that during phase D most (i.e. $\sim$ 60 \%) of the total
%unabsorbed flux is due to the thermal component.

However, the above results do not necessarily
imply that the soft component is pulsed. In
fact an acceptable fit ($\chi^{2}_{\nu}$/d.o.f. = 1.097/1989) can also be obtained
by imposing that the  black--body component parameters
be the same in the four spectra, as shown in Tab~\ref{4spectra_BBcommon}.

\begin{table}[htbp]
\caption{Best--fit values for the power--law parameters, when the four spectra are fitted simultaneously with common values of N$_{\rm H}$ ($(6.1^{+0.3}_{-0.1})\times10^{21}$ cm$^{-2}$), kT$_{\rm BB}$ ($1.40^{+0.02}_{-0.04}$ keV) and R$_{\rm BB}$ ($111^{+1}_{-4}$ m). Errors are at 90 \% confidence level for a single interesting parameter}\label{4spectra_BBcommon}
\begin{tabular}{c|cccc} \hline
Spectral            &   \multicolumn{4}{c}{Phase Interval}                                          \\
Parameter           &   A           &   B           &   C           &   D           \\ \hline
$\Gamma$   &   $1.98^{+0.03}_{-0.04}$  &   $1.57^{+0.02}_{-0.02}$  &   $1.56^{+0.04}_{-0.03}$   &   $2.63^{+0.13}_{-0.08}$  \\
I$_{\rm PL}^{a}$    &   $2.40^{+0.06}_{-0.06}$  &   $3.47^{+0.05}_{-0.04}$   &   $1.39^{+0.03}_{-0.05}$  &   $0.97^{+0.04}_{-0.05}$  \\
$f_{\rm TOT}^{b}$       &   2.08        &   3.43        &   1.81        &   1.19        \\
$f_{\rm PL}^{b}$        &   1.36    &   2.71    &   1.09    &   0.47     \\
 & (65 \%) & (79 \%) & (60 \%) & (40 \%) \\
$f_{\rm BB}^{b}$        &   0.72    &   0.72    &   0.72    &   0.72     \\
 & (35 \%) & (21 \%) & (40 \%) & (60 \%) \\ \hline
\end{tabular}
\begin{small}
\\
$^{a}$ Intensity of the power--law component in units of $10^{-3}$ ph cm$^{-2}$ s$^{-1}$ keV$^{-1}$ at 1 keV

$^{b}$ Unabsorbed flux in the energy range 0.3--10 keV, in units of $10^{-11}$ erg cm$^{-2}$ s$^{-1}$
\end{small}
\end{table}

\begin{figure}[h]
%\tabcapfont
%\begin{tabular}{c@{\hspace{1pc}}c}
\centering
\resizebox{\hsize}{!}{\includegraphics[angle=-90]{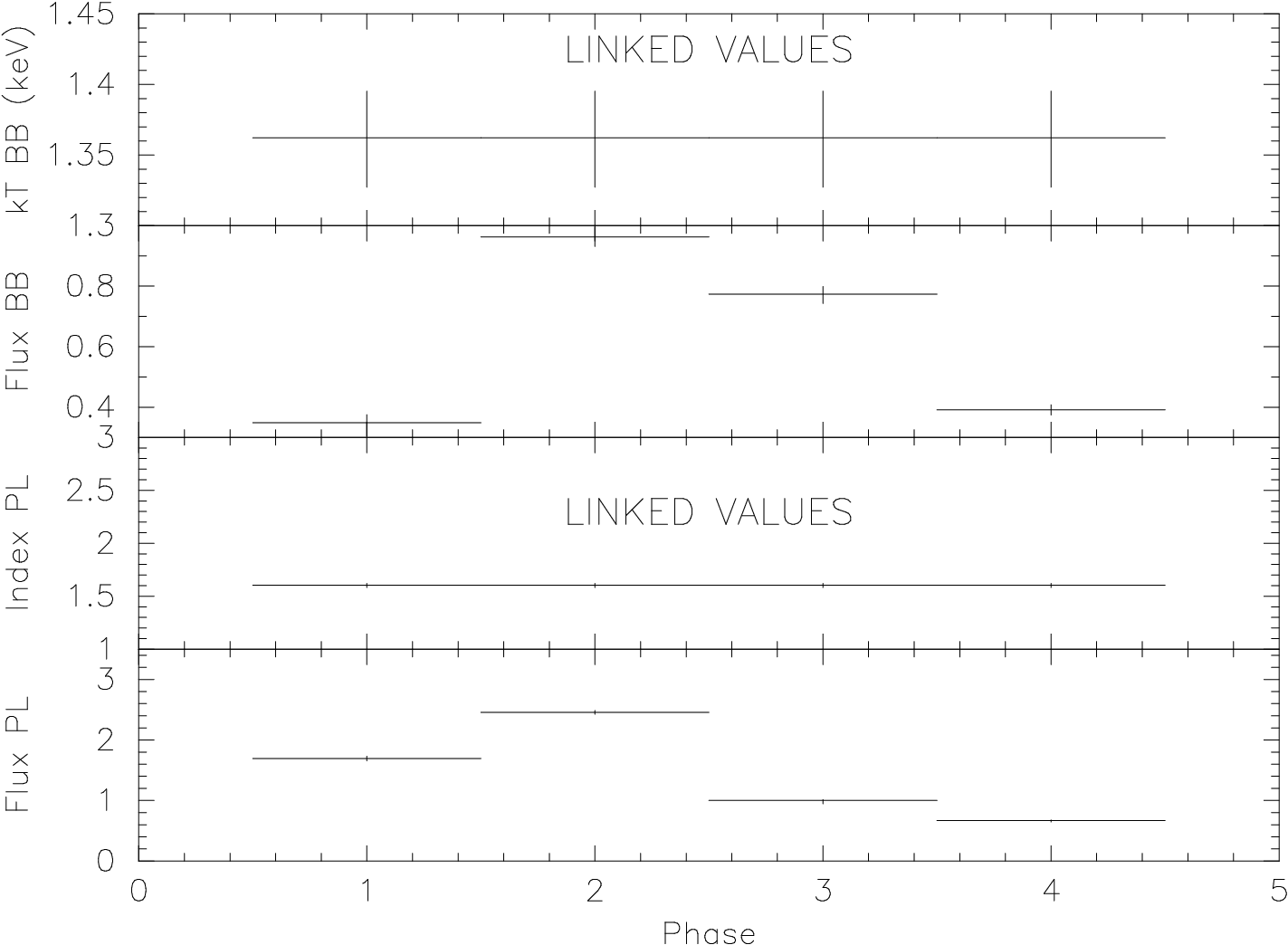}}
\resizebox{\hsize}{!}{\includegraphics[angle=-90]{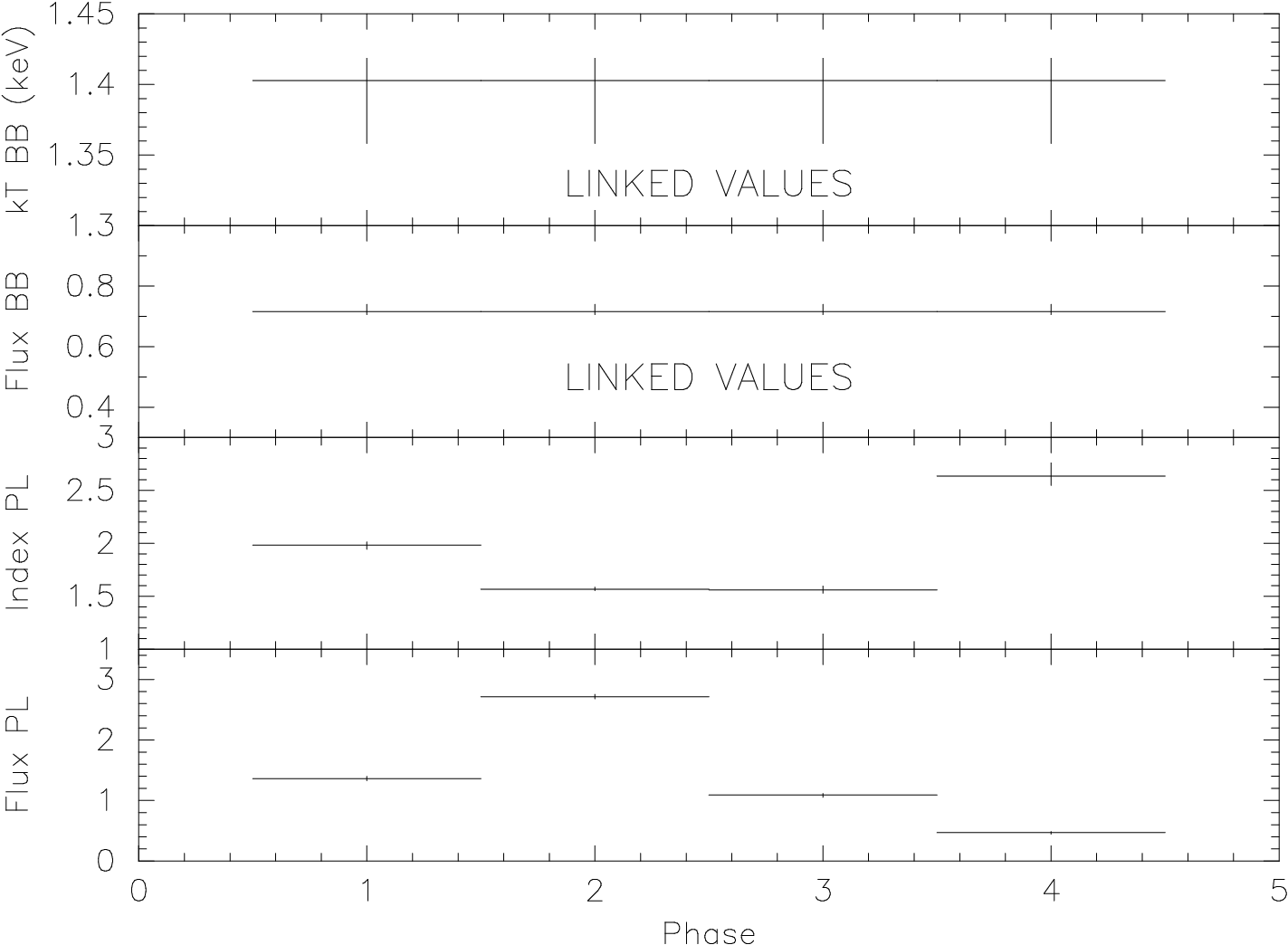}}
%\end{tabular}
\caption{Pulse--phase dependence of the black--body temperature, the power--law photon index and the unabsorbed flux of the two components (in units of 10$^{-11}$ erg cm$^{-2}$ s$^{-1}$), in the case of common temperature and index (\textit{top}) and of common black--body (\textit{bottom})}\label{both_4spectra}
\end{figure}

The above results are summarized in Fig.~\ref{both_4spectra},
where we show the pulse--phase dependence of the black--body
temperature, the power--law photon index and the unabsorbed flux
of the two components for both fits. Since they have a similar statistical
quality, we can neither confirm nor deny that the thermal
component is variable.

\section{Discussion}\label{sec:6}

\begin{figure}[h]
\centering
\resizebox{\hsize}{!}{\includegraphics[angle=-90,clip=true]{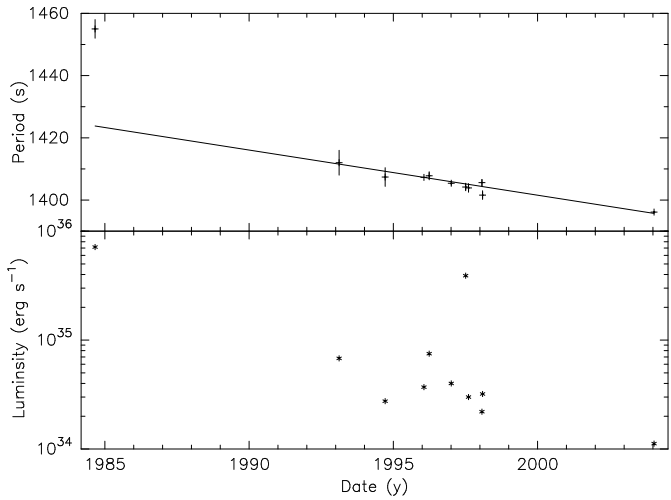}}
\caption{Pulse period and luminosity history of \RX~from 1984 August 27 to 2004 January 15. The solid line in the upper panel is the best--fit linear model of the reported values and has a slope ${\rm \dot P}= -4.6\times10^{-8}$ s s$^{-1}$. The luminosities of the lower panel are based on the unabsorbed flux in the 2--10 keV energy range and on a source distance of 2.5 kpc.}\label{luminosity}
\end{figure}

The \XMM~data described here were obtained almost 6 years after
the latest X--ray observation of \RX, which was performed by
\SAX~on 1998 February 3 \citep{Mereghetti+00}.
%: no other
%observations were performed in the meanwhile, while several data
%were accumulated in the previous years also with
%\EXOSAT~\citep{White+87}, \ROSAT~\citep{Hellier94},
%\ASCA~\citep{Haberl+98a} and \XTE~\citep{Haberl+98b}.
Therefore it is interesting to compare our results with those
obtained in the past.
%
%From the timing point--of--view,  the comparison of the measured
%period with the corresponding values obtained by the previous
%observations allows to constrain the source spin--up.
In Fig.~\ref{luminosity} we show the long term evolution of the
source spin period and luminosity since the time of the
\EXOSAT~discovery in 1984. If we exclude the first observation,
when the source was in outburst, a linear fit of all the periods
gives a spin--up at an average rate of ${\rm \dot P} =
-(4.6^{+0.1}_{-0.2})\times10^{-8}$ s s$^{-1}$, similar to that
measured until 1998. This result suggests that during the 6 years
between the \SAX~and the \XMM~observations, the momentum transfer
onto the neutron star has proceeded with no major changes.
%This is also consistent with the source luminosity history.
%: there we report the
%source luminosity values based on the unabsorbed fluxes in the
%2--10 keV energy range and assuming  a  distance of 2.5 kpc.
The flux  detected by \XMM~corresponds to a luminosity of
$\sim1.5\times10^{34}$ erg s$^{-1}$,  a factor $\sim$ 2 lower than
the minimum level observed in the previous years, indicating that,
after the outburst of July 1997 \citep{Haberl+98b}, \RX~ has been
continuously fading.

% It is interesting to note that \RX~was observed by \XMM~at
% a lower flux level than in the previous cases: the unabsorbed flux
% in the 2--10 keV energy band is $\sim1.5\times10^{-11}$ erg
% cm$^{-2}$ s$^{-1}$, instead than $\sim3\times10^{-11}$ erg
% cm$^{-2}$ s$^{-1}$ measured by \SAX; in the 2--20 keV range the
% corresponding values are $\sim3\times10^{-11}$ erg cm$^{-2}$
% s$^{-1}$ for \XMM~and $\sim1.4\times10^{-10}$ erg cm$^{-2}$
% s$^{-1}$ for \XTE.

Comparison with the \XTE~results reported in Fig.2 of
\citet{Mereghetti+00} shows that, despite the luminosity
variation, the shape of the pulse profiles in the energy interval
common to the two instruments (2-10 keV) has not changed. The
\XMM\ data also confirm that, above 2 keV, there is a
%The new data indicate that also in the softest range
%0.3--2 keV, not observed with \XTE, the source has a comparable
%variability pattern.
%
%The comparison of these data with those reported in Fig.4 of
%\citet{Mereghetti+00} suggests a few comments, also thanks to the
%extension of the sampled energy range below 2 keV: This figure
%clearly shows the spectral variations, and in particular that:
%
%\begin{itemize}%
%
%item  there is a
significant spectral softening in the initial rising part of the
pulse (phase interval A).

%\item contrary to the previous findings, it seems that during
%phase interval C there is a spectral hardening;

%\item
%during phase interval D the spectrum is harder than the average
%one above 2 keV, but below 1 keV there is also a significant
%excess that was not visible with \XTE

%\end{itemize}

The main advantage of the data reported here, with respect to
previous observations of this source, is the better coverage of
the energy range below 2 keV.
%The high sensitivity of EPIC in this range
%Our data cover the energy range 0.3--10 keV, while those reported
%by \citet{Mereghetti+00} span between 2 and 20 keV. Therefore in
%our analysis we can study better the soft end of the source
%spectrum but miss the hard part. On the other hand, thanks to the
%narrow \textit{Point Spread Function} of the \XMM~mirrors, we
%could consider the spectrum of \RX~with no contamination from
%other nearby sources. If we consider the spectrum of the whole
%observation, i
In comparison with the \SAX~results, the phase averaged
spectroscopy gives a smaller hydrogen column density: N$_{\rm
H}=(5.09^{+0.24}_{-0.23})\times10^{21}$ cm$^{-2}$ instead of
(1.2$\pm$0.3)$\times10^{22}$ cm$^{-2}$. This value is in better agreement
with that expected from the optical observations which give
$E(B-V)$ = 0.93 \citep{Reig+97}. Assuming A$_{\rm V}$ = 3.1 $E(B-V)$ and the average relation A$_{\rm V}$ = N$_{\rm H} \times 5.59 \times 10^{-22}$ cm$^{-2}$ between
optical extinction and X-ray absorption \citep{PredehlSchmitt95}, this would
predict N$_{\rm H}=5.16\times10^{21}$ cm$^{-2}$.

% and also a smaller photon--index
%($\Gamma=1.34^{+0.05}_{-0.06}$ instead than 1.67$\pm$0.10).

Moreover, we find evidence of a thermal component which was never
observed before in this source.
%
% Finally, the phase--resolved spectroscopy described in
% Sec.~\ref{spectroscopy} yielded two interesting results: on one
% hand, in the corresponding phase ranges we obtained lower
% best--fit values for both the hydrogen column density and the
% photon--index; on the other hand, the variation of both these
% parameters with the phase interval is the same observed by \XTE.
%
%In reference to the soft excess,
%
In the last years, similar soft excesses have been detected in
several high mass X--ray binary pulsars. The properties of these
sources are summarized in Table~\ref{pulsars}.
% we summarize the main
%characteristics of those pulsars in which a soft excess has been
%detected.
%
%These observations have been recently reviewed by who considered
%several possible explanations for their origin.
%have investigated
%the brightest of these sources in order to explain its physical origin.
\citet{Hickox+04} showed that a soft excess has been detected in
all the X-ray pulsars with a sufficiently high flux and small
absorption. In fact, most of the soft excess sources are at small
distances and/or away from the Galactic plane (most of them are
in the Magellanic Clouds). This suggests that the presence of a
soft spectral component could be a very common, if not an
ubiquitous, feature intrinsic to X--ray pulsars.
%, which can be detected in sources which are characterized by both
%a high flux and a low interstellar absorption. .Moreover, they have found
These authors concluded that the origin of the soft component is
related to the source total luminosity. When $L_{\rm
X}\ge10^{38}$ erg s$^{-1}$ the luminosity and the shape
of the soft component can be explained only by reprocessing of
hard X--rays from the neutron star by optically thick accreting
material, most likely near the inner edge of the accretion disk.
In  less luminous sources, with $L_{\rm X}\le10^{36}$ erg
s$^{-1}$, the soft excess can be due to other processes,
such as emission by photo--ionized or collisionally heated diffuse
gas or thermal emission from the surface of the neutron star.
Finally, in the sources of intermediate luminosity, either or both
of these types of emission can be present.

\begin{figure}[h]
%\begin{tabular}{c@{\hspace{1pc}}c}
\centering
\resizebox{\hsize}{!}{\includegraphics[angle=-90]{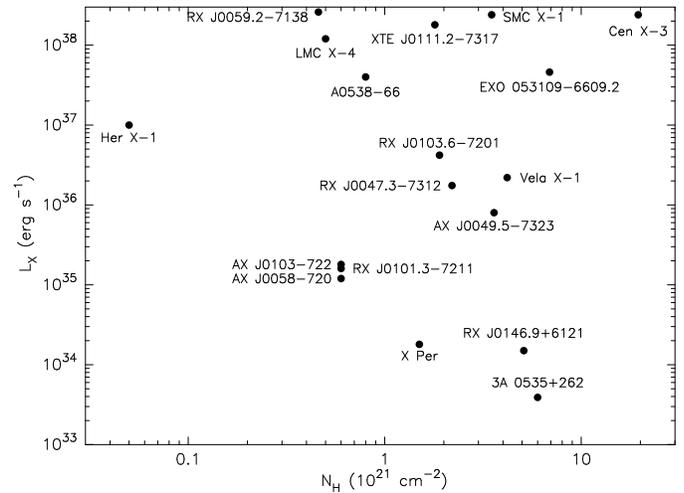}}
%\resizebox{\hsize}{!}{\includegraphics[height=8cm,angle=-90]{FX_NH.ps}}
%\end{tabular}
\caption{Total X--ray luminosity  of the sources of
Tab.~\ref{pulsars} as a function of the interstellar
absorption.}\label{luminosities}
\end{figure}

In Fig.~\ref{luminosities} we have plotted the X--ray luminosity
versus interstellar absorption of the pulsars with a soft excess.
We can see that \RX, together with X Persei and 3A 0535+262, is at least one
order of magnitude less luminous  than  all the other  pulsars
displaying a soft excess. We note that the ratio between the
unabsorbed fluxes of the thermal and the power--law components in
\RX\ agrees well with the average value measured in the other
sources. However, unlike some of the other sources, the soft excess
could be fitted only with a black--body component, while any other
emission model was rejected.

\begin{landscape}
\begin{table}
\caption{Orbital and spectral parameters for XBPs with a detected soft excess (SE).}\label{pulsars}
\begin{scriptsize}
\begin{tabular}{cccccccccccc} \hline
Source$^{a}$        & Location & Distance   & Companion     & P$_{\rm pulse}$ & L$_{\rm X}^{b}$                 & Flux              & N$_{\rm H}^{c}$   & L$_{\rm SE}$/L$_{\rm X}$ & SE model$^{d}$ & kT$_{\rm BB}$ & SE Pulses$^{e}$ \\
            &      & (kpc)  & Star      & (s)         & (ergs s$^{-1}$, keV)            & (ergs cm$^{-2}$ s$^{-1}$) & ($10^{21}$ cm$^{-2}$) &            &          & (keV)     &       \\ \hline
Her X--1$^{1}$      & Galaxy   & $\sim$5    & A7 V      & 1.24        & $1.0\times10^{37}$ (0.3-10)         & $3.3\times10^{-9}$        & 0.05          & 0.04--0.10         & BB, BB+LE        & 0.09--0.12    & Yes       \\
SMC X--1$^{2}$      & SMC      & 65     & B0 Ib     & 0.7         & $2.4\times10^{38}$ (0.7--10)        & $4.7\times10^{-10}$       & 2--5          & 0.036          & BB, TB, SPL      & 0.15--0.18    & Yes       \\
LMC X--4$^{3}$      & LMC      & 50     & O7 III--O IV  & 13.5        & $1.2\times10^{38}$ (0.7--10)        & $4.0\times10^{-10}$       & $\sim$0.5     & 0.064          & BB, BB+TB, COM, SPL  & 0.15      & Yes       \\
\xtej$^{4}$         & SMC      & 65     & B1 IVe    & 30.95       & $1.8\times10^{38}$ (0.7--10)        & $3.6\times10^{-10}$       & 1.8           & $\sim$0.10         & SPL          & \nodata   & Yes       \\
\rxj$^{5}$      & SMC      & 65     & B1 IIIe   & 2.76        & $2.6\times10^{38}$ (0.1--10)        & $5.1\times10^{-10}$       & 0.42--0.50        & 0.31           & MEK, SPL     & \nodata   & No        \\
\fouru$^{6}$        & Galaxy   & ?      & Low--mass     & 7.7         & $2.6\times10^{34}D_{\rm kpc}^2$ (0.5-10)    & $2.2\times10^{-10}$       & 0.6           & 0.10           & BB           & 0.34      & No        \\
Cen X--3$^{7}$      & Galaxy   & $\sim$8    & O6--O8 II     & 4.8         & 2.4$\times10^{38}$ (0.1--10)        & 3.2$\times10^{-8}$        & 19.5          & $\sim$0.7      & BB           & 0.11      & Yes       \\
Vela X--1$^{8}$     & Galaxy   & 1.9    & B0.5 Ib   & 283         & $2.2\times10^{36}$ (2--10)          & $5.1\times10^{-9}$        & 4.2           & $\sim$0.01         & TB           & \nodata   & No        \\
X Per$^{9}$         & Galaxy   & 0.95   & O9.5pe    & 837         & $1.8\times10^{34}$ (0.3--10)        & $1.7\times10^{-10}$       & 1.5           & 0.24           & BB           & 1.45      & ?     \\
\exo$^{10}$         & LMC      & 50     & B0.7 Ve   & 13.7        & $4.6\times10^{37}$ (0.2--10)        & $1.5\times10^{-10}$       & 6.9           & ?          & MEK+PL       & \nodata   & Yes       \\
\aofive$^{11}$      & LMC      & 50     & B2 IIIe   & 0.069       & $4.0\times10^{37}$ (0.1--2.4)       & $1.3\times10^{-10}$       & 0.8           & ?          & BB, TB       & $\sim$0.2 & ?     \\
RX J0047.3-7312$^{12}$  & SMC      & 65     & B2e       & 263         & $1.5,2\times10^{36}$ (0.7--10)      & $3.0,4.0\times10^{-12}$   & 0.96,3.4      & 0.03--0.09,0.68    & BB           & 0.6, 2.2  & Yes       \\
RX J0101.3-7211$^{13}$  & SMC      & 65     & Be        & 452         & $1.6\times10^{35}$ (0.3--10)        & $3.2\times10^{-13}$       & 0.6           & ?          & MEK          & \nodata   & ?     \\
RX J0103.6-7201$^{14}$  & SMC      & 65     & O5 Ve     & 1323        & $0.8-7.5\times10^{36}$ (0.2--10)        & $1.6-14.8\times10^{-12}$  & 1.9           & ?          & MEK          & \nodata   & No        \\
AX J0049.5-7323$^{15}$  & SMC      & 65     & B2 Ve     & 751         & $7-9\times10^{35}$ (0.2--10)        & $1.4-1.8\times10^{-12}$   & 3.6           & ?          & ?            & \nodata   & Yes       \\
AX J0058-720$^{13}$     & SMC      & 65     & Be        & 281         & $1.2\times10^{35}$ (0.3--10)        & $2.4\times10^{-13}$       & 0.6           & ?          & ?            & \nodata   & Yes       \\
AX J0103-722$^{13}$     & SMC      & 65     & B0 IV     & 342         & $1.8\times10^{35}$ (0.3--10)        & $3.6\times10^{-13}$       & 0.6           & ?          & MEK          & \nodata   & ?     \\
3A 0535+262$^{16}$	& Galaxy	& 2.0	& O9.7 IIIe	& 103.4	& $3.9\times10^{33}$ (2--10)	& $8.2\times10^{-12}$	& 6.0	& 0.35	& BB	& 1.33	& ?	\\
\RX         & Galaxy   & 2.5    & B0 IIIe   & 1395        & $1.5\times10^{34}$ (0.3--10)        & $2.0\times10^{-11}$       & 5.1           & 0.25           & BB           & 1.11      & ?     \\ \hline
\end{tabular}
\end{scriptsize}
\\
$^{a}$ References: (1) \citet{dalFiume+98}, \citet{Endo+00}, \citet{Ramsay+02}; (2) \citet{Woo+95}, \citet{Paul+02}; (3) \citet{Woo+96}, \citet{LaBarbera+01}, \citet{Naik&Paul04}; (4) \citet{Yokogawa+00a}; (5) \citet{Kohno+00}; (6) \citet{Schulz+01}; (7) \citet{Burderi+00}; (8) \citet{Haberl94}, \citet{Orlandini+98a}, \citet{Kreykenbohm+02}; (9) \citet{diSalvo+98}, \citet{Coburn+01}; (10) \citet{Haberl+03}; (11) \citet{MavroHaberl93}; (12) The two set of values are base on \citet{Ueno+04} and \citet{Majid+04}, respectively; (13) \citet{Sasaki+03}; (14)\citet{Sasaki+03}, \citet{Haberl&Pietsch05}; (15) \citet{Yokogawa+00b}, \citet{Haberl&Pietsch04}; (16) \citet{MukherjeePaul05}.
\\
$^{b}$ For each source also the reference energy range is reported.
\\
$^{c}$ The reported values refer only to the interstellar absorption, not to the source intrinsic absorption.
\\
$^{d}$ Spectral models used for the soft excess are: BB = blackbody; TB = thermal bremsstrahlung; SPL = soft power--law or broken power--law; MEK = MEKAL thin thermal model; COM = Comptonization model; LE = broad low--energy line emission. Commas indicate separate fits, plus signs indicate fits with two components.
\\
$^{e}$ Pulsation of the emission component that traces the soft excess.
\end{table}
\end{landscape}

Based on the results obtained by \citet{Hickox+04}, the luminosity
of \RX\ is too small for an interpretation of the soft excess in
terms of reprocessing of the hard X--ray photons in optically
thick accreting material. Moreover, both the thermal emission and
the hard X--ray reprocessing in a diffuse, optically thin gas
around the neutron star are unlikely, since these processes would
not give a black--body spectrum. We therefore favor the
interpretation of the soft excess in \RX\ as thermal emission from
the neutron star polar cap. If we assume that the source is in the `accretor' status, with matter accretion on the NS surface, the blackbody emitting radius of $\sim$ 140 m
is consistent with the expected size of the polar cap.
In fact, if M$_{\rm NS}$ = 1.4 M$_{\odot}$ and R$_{\rm NS}$ = $10^6$ cm, the source luminosity of $\sim$ $10^{34}$ erg s$^{-1}$ implies an accretion rate $\dot M \sim 5 \times 10^{13}$ g s$^{-1}$ and, adopting B$_{\rm NS} = 10^{12}$ G, a magnetospheric radius R$_m \simeq 2.4 \times 10^9$ cm \citep{Campana+98}. In this case, based on the relation R$_{col} \sim$ R$_{\rm NS}$ (R$_{\rm NS}$/R$_m$)$^{0.5}$ \citep{Hickox+04}, we would obtain R$_{col} \sim$ 200 m.
%On the other hand, its high temperature and small emission
%radius suggest that the origin of this soft excess is in the
%neutron star polar caps.
If this description is correct, we would expect to observe some
variability of the soft component along the pulse phase. From this
point of view the phase--resolved spectroscopy provides no
conclusive results, since it proves that both a variable and a
constant thermal component can account for the observed spectral
variability.

We finally note that \RX~is in many respects very similar to X
Per and 3A 0535+262, which are very low luminosity Be/NS binary with a
long pulse period. The soft excesses of these
sources have similar properties, since they tend to have a higher temperature ($>$ 1 keV) and a smaller emission radius ($\sim$ 0.1 km) compared to the soft excesses observed in high luminosity systems, which have a temperature of about 0.1 keV and an emission radius
of a few hundred km. In these three low luminosity systems the soft excess contributes for 25--35 \% of the total flux \citep{Coburn+01,MukherjeePaul05}.
Also in the case of X Per and 3A 0535+262 this excess has been attributed to the
emission from the polar caps.
%,
%therefore we infer that the same description could be proposed
%also for \RX.

%In  Tab.~\ref{pulsars} we summarize the main characteristics of
%those pulsars in which a soft excess has been detected.
%Comparison with the results obtained for \RX~provides a few
%interesting information:

%\begin{itemize}

%\vspace{-3mm}

%The above items show that, although \RX~is a very low luminosity
%pulsar in the galactic plane, it is very near and has little
%absorption: therefore it is characterized by a high low energy
%flux, comparable to the one observed in the other pulsars which
%show a soft excess. Hence the detection of the same spectral
%feature in \RX~extends the low luminosity end of these type of
%sources, as shown in Fig.~\ref{luminosities}.

% They show that, in all the four phase intervals, both N$_{\rm H}$
% and $\Gamma_{\rm PL}$ have smaller values than in the
%corresponding \XTE~measurements. However, it is interesting to
%note that, for both of them, the variation pattern between the
%various phases is the same observed by \XTE: the photon index
%$\Gamma_{\rm PL}$ continuously decreases from phase intervals A to
%D; the absorption column N$_{\rm H}$ first increases from A to B
%and then decreases from B to D. The differences in the blackbody
%temperatures between the phase intervals A--B and C--D are always
%within the estimated errors. In all the phase intervals the flux
%raction due to the thermal component is $\sim$ 25\%, i.e.
%%comparable to the percentage measured in the total spectrum; only
%in the case of phase interval C it increases up to $\sim$ 40\%.

\section{Conclusions}

We have reported on the analysis of the data collected by \XMM~in
a $\sim$ 42 ks observation of the Be/neutron--star X--ray pulsar
\RX.

The unabsorbed flux corresponds to a source luminosity L$_{\rm
X}\sim1\times10^{34}$ erg s$^{-1}$ in the 2--10 keV energy range,
about 50~\% smaller than the lowest level ever observed from this
source, indicating a monotonic source fading over long
time scales.

Thanks to the high effective area of \XMM~also at low energies,
we could perform the first accurate spectral study below 2 keV for
this source. In the phase--averaged spectrum we have revealed the
presence of a significant soft excess over the primary power--law
component: this excess can be described by a black--body with
kT$_{\rm BB}\sim$ 1 keV, while any attempt to fit it with a
different emission model was unsuccessful.

The phase--resolved spectroscopy has confirmed the  large spectral
variability along the pulse period already observed above 2 keV.
Unfortunately, with the current data, it is not possible to
derive compelling results on the phase variability of the soft
excess component. Although the emission below 2 keV is clearly
pulsed and the low energy part of the spectrum varies with the
phase, we have shown that such variations can be explained equally
well by changes in the blackbody component or in the power law
component alone.

Clearly the relatively small distance and low interstellar
absorption toward \RX\ played a role in the possibility of
detecting a soft excess in such a low luminosity pulsar.
Comparison with other X--ray binary pulsars shows that, so far,
the soft excess had been detected only in much brighter sources.
The data reported here support the hypothesis that a soft thermal
component is an ubiquitous emission feature of this class of
sources. In this sense, it would be very interesting to use the
large collecting area of \XMM~in long observations of the faintest
and longest period Be binaries, both in the SMC and in the Milky
Way, such as, for example, the persistent low luminosity systems
RX J0440.9+4431 and RX J1037.5-5647 \citep{Reig&Roche99}.

\begin{acknowledgements}
We thank A. Tiengo for his useful suggestions. The {\em XMM--Newton} data analysis is supported by the Italian Space Agency (ASI), through contract ASI/INAF I/023/05/0.
\end{acknowledgements}

\bibliographystyle{aa}
\bibliography{biblio}

\end{document}